\documentclass{interact}

\usepackage{subcaption}
\usepackage{booktabs}
\usepackage{hyperref}

\usepackage[numbers,sort&compress]{natbib}%
\bibpunct[, ]{[}{]}{,}{n}{,}{,}%
\makeatletter%
\def\NAT@def@citea{\def\@citea{\NAT@separator}}%
\makeatother%

\usepackage{letltxmacro}

\newcommand{\Fl}{\ensuremath{\mathrm{F}}}
\newcommand{\Rel}{\ensuremath{\mathrm{Re}}}
\newcommand{\Lhat}{\ensuremath{{\hat{L}}}}
\newcommand{\Uhat}{\ensuremath{{\hat{U}}}}

\newcommand{\khkz}{\ensuremath{(\kappa_h, \kappa_z)}}
\newcommand{\kz}{\ensuremath{\kappa_z}}
\newcommand{\kh}{\ensuremath{\kappa_h}}

\begin{document}

\title{
Unsupervised Machine Learning to Teach Fluid Dynamicists to Think in 15 Dimensions}

\author{
  \name{Stephen\ M.\ de Bruyn Kops\textsuperscript{a},\thanks{Corresponding author: S. M. de Bruyn
      Kops at debk@umass.edu}
Daniel\ J.\ Saunders\textsuperscript{b,c},
Edward\ A.\ Rietman\textsuperscript{b},
Gavin\ D.\ Portwood\textsuperscript{a,d}}
  \affil{
\textsuperscript{a}Department of Mechanical and Industrial Engineering,
University of
Massachusetts Amherst, Amherst, Massachusetts, USA;
\textsuperscript{b}College of Computer and Information Sciences, University of
Massachusetts Amherst, Amherst, Massachusetts, USA;
\textsuperscript{c}Fomoro AI, San Francisco, CA, USA;
\textsuperscript{d}Methods and Algorithms (XCP-4), Computational Physics Division, 
       Los Alamos National Laboratory, Los Alamos, New Mexico, USA
}}

\maketitle
\begin{abstract}
An autoencoder is used to compress and then reconstruct three-dimensional
stratified turbulence data in order to better understand fluid dynamics by studying the errors in the reconstruction. 
The original single data set is resolved on approximately $6.9\times10^{10}$ grid points, and 15 fluid variables in three spatial dimensions are used, for a total of about $10^{12}$ input quantities in three dimensions.  The objective is to understand which of the input variables contains the most relevant information about the local turbulence regimes in stably stratified turbulence (SST). This is accomplished by observing flow features that appear in one input variable but then `bleed over' to multiple output variables.  The bleed over is shown to be robust with respect to the number of layers in the autoencoder.  In this proof of concept, the errors in the reconstruction include information about the spatial variation of vertical velocity in most of the components of the reconstructed rate-of-strain tensor and density gradient, which suggests that vertical velocity is an important marker for turbulence features of interest in SST.  This result is consistent with what fluid dynamicists already understand about SST and, therefore, suggests an approach to understanding turbulence based on more detailed analyses of the reconstruction on errors in an autoencoding algorithm.   
\end{abstract}


\section{Introduction}
Stably stratified turbulence (SST) is a model flow that is potentially useful
for understanding portions of the
deep ocean, stratosphere, and atmospheric boundary layer as well as large refrigerated storage facilities and other engineered systems. In particular SST
describes the relatively small-scale dynamics, in length and time, at which
turbulence and mixing occurs.  These small scales can be profoundly important
for, say, the conversion of kinetic energy to heat in the oceans or the drag
on a submerged object, but they are typically impractical to resolve in
simulations of geophysical flows, or even in simulations of engineering flows.
Consequently, it is necessary to parametrize the effects of these small scales
in terms of the larger scales.  As detailed later in this introduction,
machine learning is advancing in tuning parametrizations once they have been
conceived by humans.  In this paper we report on advances that go beyond
existing approaches in order to use machine
learning to teach fluid dynamicists the basis for the parametrization in parameter spaces
with too many dimensions for people to easily work with.

The objective of parameterizing small scale effects is common to all
turbulence modeling, but SST is complicated by the fact that turbulence
appears to occur in several dynamically distinct regimes.  It is widely
hypothesized that effective models will begin with correctly identifying the
turbulence regime that dominates some region in space.  Portwood et
al.\cite{portwood16} use cumulative filtered distribution functions to
distinguish three such regimes in direct numerical simulations (DNSs) of SST
resolved on up to $8192 \times 8192 \times 1024$ grid points.  Those results
are discussed in more detail later in the introduction, but for the moment we
simply observe that the technique used for that research, while proven
valuable, is extremely laborious and also subjective in that it is predicated
on there being three preconceived subregimes in the flows.  The motivation of
the research reported here is to develop a machine learning technique
that can consider simultaneously many flow variables in three spatial
dimensions at $O(10^{11})$ points in space to identify dynamically distinct
flow regimes, assuming they exist, without {\em apriori} assumptions about
those regimes.

The goal of this project is to determine whether an autoencoder can reveal which of 15 flow variables contains information about the flow regimes of interest. This is done by reconstructing a very large snapshot in time of a simulated fluid flow and observing the `bleed over' from a three-dimensional input variable into other three-dimensional output variables.  In other words, the objective is not to generate a perfect reconstruction, or to induce a machine to learn the statistical parameters describing a population of flows, either of which might be approached with sufficient training samples, but rather to derive physical insight from imperfections in the reconstruction for a single but very large sample representing a fluid flow field at an instant in time.  Given that the largest fluid turbulence simulations to date are run on about $4\times10^{12}$ grid points, our test with $6.9\times10^{10}$ grid points and 15 flow variables ($10^{12}$ total samples) makes this the first, to our knowledge, almost-full-scale test of machine learning applied to fluid turbulence.

\subsection{Dynamically distinct regimes in SST}

A flow is stably stratified by when fluid with lower density sits on top of
fluid with higher density.  This occurs due to variations in
temperature in any fluid, salt concentrations in water, and humidity levels in
air.  When a flow is stably stratified, vertical motion is constrained by
buoyancy so that it is anisotropic at large length scales but may be
approximately isotropic at smaller length scales that are not strongly
affected by buoyancy.  Among the earliest research in SST is that of Lin and
Pao who observed that pancake-like structures form due to the stabilizing
effect of buoyancy \cite{lin79}.
Lilly \cite{lilly83} proposed that these pancakes shear against each and that
turbulence errupts because the thickness of
the pancakes dynamically adjusts so that they are susceptible to shear
instabilities; this hypothesis was formalized mathematically by Billant
and Chomaz \cite{billant01}.  Riley and Lindborg \cite{riley08} review the
main concepts of SST and de Bruyn Kops and Riley \cite{debk19} review many
of the laboratory experiments and direct numerical simulations in SST.

\begin{figure}
  \begin{center}
    \includegraphics{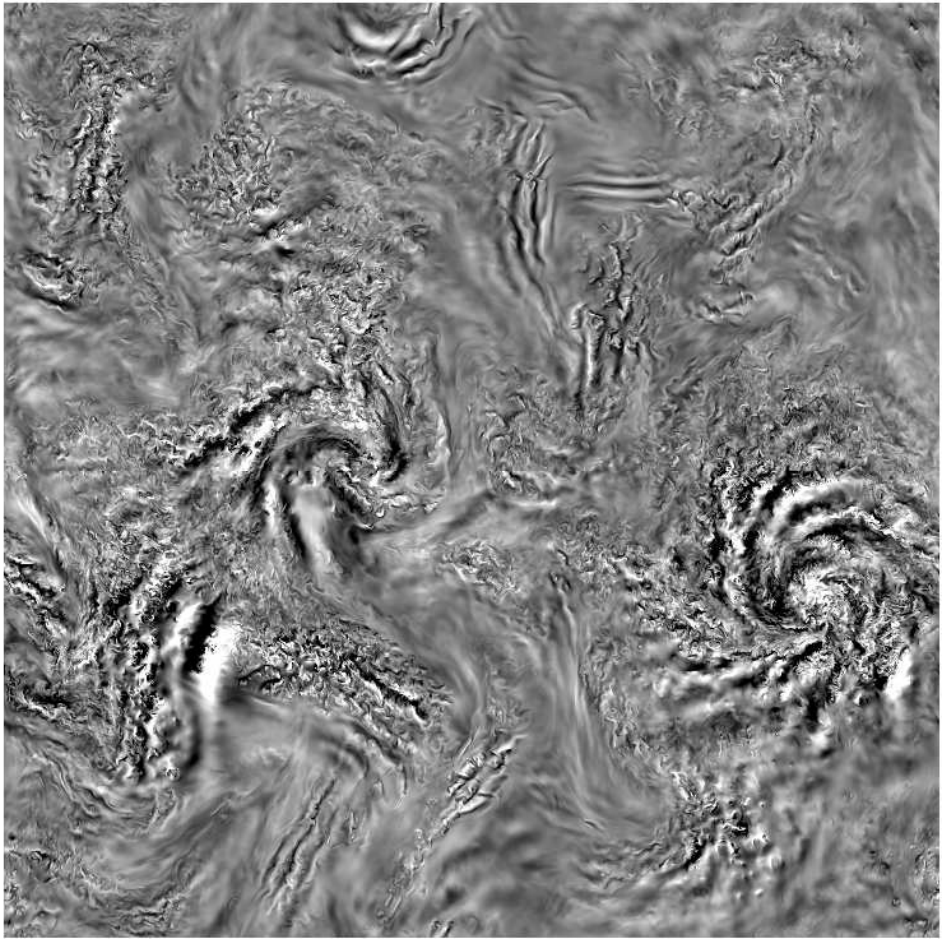}
  \end{center}
  \caption{
    Vertical velocity on a horizontal plane in stratified turbulence
    normalized by the RMS value.  White and black indicate upward and downward
    velocity respectively.  The simulation is case F3 in \cite{almalkie12a}. The simulation domain is $8192\times8192\times1024$ grid points.
\label{fig1}}
\end{figure}

In Figure~\ref{fig1}
is shown the vertical velocity on a horizontal plane in
simulated SST.  These data are from case F3 in Almalkie and de Bruyn Kops
\cite{almalkie12a} and are qualitatively consistent with those 
from simulations reported in, e.g.,
\cite{riley03,brethouwer07,waite11,kimura12,bartello13,maffioli16,debk19}.  Of
interest in the current context is that it appears that different
types of turbulence may be ocurring in the flow.  For example, on the right
side of the figure is what appears to be an intense patch of turbulence
whereas above it on the same side is a region that appears to be almost
quiescent.  Elsewhere in the image, one might observe regions that seem
qualitatively different from these two.

\begin{figure}
  \begin{center}
    \includegraphics{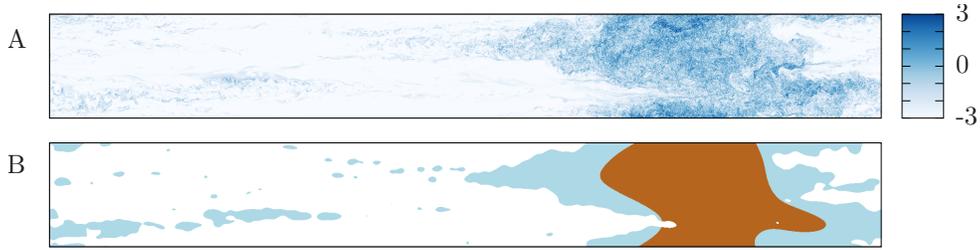}
  \end{center}
  \caption{
    A: Normalized common logarithm of the potential enstrophy on a vertical slice through
    simulated SST.  B: The same slice with the turbulence regimes classified
    by the method of Portwood et al.\cite{portwood16} with white indicating
    quiescent flow, light blue indicating layered turbulence, and brown
    indicating space-filling turbulence.
\label{fig2}}
\end{figure}

Based on the foregoing observation coupled with more detailed analyses,
Portwood et al.\ hypothesized that SST can be treated as amalagmation of
dynamically distinct flow regimes.  They examined the potential enstrophy, which is the
square of the vorticity component aligned with the local density gradient.
This quantity is well-established for distinguishing turbulence from
non-turbulence \cite{corrsin55,watanabe16}.  A plot of potential enstrophy from case F3 in
\cite{almalkie12a} is shown in panel A of Figure~\ref{fig2}.
From this figure,
it is evident that that some of the turbulence is space filling, that is, it
extends the vertical extent of the domain, while some is layered.  Recall that
Lilly \cite{lilly83} hypothesized that the pancake structures that form due to
the stabilizing effects of buoyancy shear against each other to cause
turbulence.  Portwood et al.~hypothesize that this mechanism explains the
horizontal layers of quasi-two-dimensional (Q2D) turbulence evident in
Figure~\ref{fig2}. They also hypothesize that when Q2D regions are sufficiently
energetic that they become fully three-dimensional and form a space-filling
patch.  Using an algorithm with a human in the loop, they classify regions in
Figure~\ref{fig2}A as being quiescent, Q2D, or space-filling patches.  These
classifications are shown in Figure~\ref{fig2}B.

Important in the context of this paper, Portwood et al.\ begin with the
framework of three dynamically distinct regimes (quiescent, layers, patches)
and then manually tune thresholding parameters to distinguish the three
regimes.  Our objective here is to determine how many dynamically
distinct regimes occur and, in the process, teach us how to distinguish them.

\subsection{Machine learning and turbulence}
The use of unsupervised methods in turbulence analysis and modeling applications has a broad historical context in applied fields of reduced-order modeling, modal decomposition and feature detection. Principal component analysis (PCA) was first applied to turbulence problems by Lumley et al.\ in the 1960's \cite{lumley67} before it found widespread popularity in signal processing and other data processing fields. PCA and other techniques (such as dynamic mode decomposition or scattering transform) aid in generation of reduced order-models where spectral representation in a reduced space can improve computation tractability of simulations and aid the interpretation of data.  An emerging alternative to these techniques is the use of an autoencoder, a type of deep learning machine. Deep learning is the study and application of multi-layered artificial neural networks (ANNs) \cite{Goodfellow-et-al-2016} applied to complex real-world problems. It has enjoyed success in computer vision, machine translation, and other fields \cite{LeCun2015DeepL}. 

Recent papers report on deep networks applied to study turbulent flow. Srinivasan et al.\ \cite{srinivasan19} explored the use of a simple multilayer perceptron (MLP) of [4500, 90, 90, 9] architecture, where the network had 4500 inputs, two hidden layers of 90 neurons each, and the output was 9 neurons. The authors also explored the use of a machine learning algorithm known as a long short-term memory (LSTM) network, which is often used to explore dynamic data, or time-series data. Their goal was to predict turbulent statistics. They used a data set of 10,000 different turbulent time series. Maulik et al.\ \cite{maulik18} also used an MLP. Their output was a classifier with the goal of identifying positive eddy-viscosity in flows in snapshot images from a database of direct numerical simulations (DNSs). 

An  autoencoder is an ANN used for unsupervised learning in which (in the `undercomplete' setting) a function to reduce the dimensionality of a particular data set is learned as a byproduct of learning to reconstruct compressed inputs \cite{Baldi:2011:AUL:3045796.3045801}. The autoencoder first encodes data into a smaller representation (code tensor) using a learned function $\mathbf{z} = e(\mathbf{x}; \theta_{e})$. The code tensor is then \textit{decoded} back into the shape of the input sample, $\hat{\mathbf{x}} = d(\mathbf{z}; \theta_{d})$. These functions are parametrized by two ANNs called the encoder and decoder networks, respectively.

Gonzalez and Balajewcz \cite{gonzalez18} use an convolutional autoencoder with a LSTM for learning low-dimensional feature dynamics of fluid systems. Their input was $128 \times 128$ snapshots. Like Mohan et al.\ \cite{mohan19} and King et al.\ \cite{king18}, they used convolutional autoencoders to `compress' the data and then, since the data represent flow dynamics, they applied a LSTM to the compressed layer. Mohan et al.\ and King et al.\ used $128 \times 128 \times 128$ image snapshots from DNS as input.  

ANNs are designed to minimize a loss function $L(\mathbf{x}, \mathbf{\hat{x}})$,  i.e., to reconstruct the input as accurately as possible. This minimization is accomplished by stochastic gradient descent (SGD) via the back-propagation algorithm. In the process of learning to reconstruct inputs, a compressed representation of the data is learned in code tensor space. Additional penalties can be added to the loss function as soft constraints on the learned representation, e.g., regularization on the network's weights (weight decay) to encourage them to be small.

One approach to using autoencoders involves minimizing the loss function $L(\mathbf{x}, \hat{\mathbf{x}})$ with respect to multiple training examples $\{\mathbf{x}_i \}_{i = 1}^{n_\textrm{train}}$.  Applications include generating reduced-order, or compressed, representations of turbulence \cite{vijayan18,king18}.  Compressed representations have been applied to temporal modeling or emulation techniques \cite{mohan19,omata19,gonzalez18} and to generate synthetic turbulence as initial conditions for simulations \cite{fukami18}. Here emulation is distinct from simulation in that the former attempts to model the latter without classical numerical techniques and at lower computational cost. Indeed, neural network models have also been developed to approach the closure problem of turbulence by tuning or generating turbulence closure in large-eddy simulation \cite{sarghini03,beck18,maulik19,nikolaou19} or Reynolds-averaged frameworks \cite{ling16,moghaddam18,zhao19}. 

Our use of an autoencoder is somewhat different from that described in the preceding paragraph. 
Like some of the investigators cited above, we use a convolutional autoencoder (CAE). Unlike those investigators our goal is to use a CAE to `discover' new features in 15-dimensional space. We are motivated by Vapnik \cite{vapnik13}:
\begin{quote}
When solving a given problem, try to avoid solving a more general problem as an intermediate step.
\end{quote}
Vapnik is saying that in most applications of machine learning the investigators are concerned with estimation of densities for a universal problem in that domain. That is, the focus is on having the learning machine induce the statistical parameters for a distribution, or on building a regression model that successfully generalizes (i.e.\ interpolates). Vapnik's argument is that even with a small data set it is possible to induce a regression or classification boundary that will be `good enough'. Our goal is not to build a general model of SST but rather to explore the use of a CAE to estimate if the learning machine can inform us about higher dimensions in stably stratified turbulence. 

Our approach is to use a very high-dimensional data set of about $10^{12}$ input quantities. We discuss a technique to down sample and reduce it to $10^8$. After reconstruction of the outputs we observe that certain features 'bleed over' from the compression layer (the code tensor). This suggests one can use a CAE to explore novel questions about code tensor and output reconstruction. 

\section{Direct Numerical Simulation Database}
The simulated flows considered in this research are solutions to the
Navier-Stokes equations in a non-rotating reference frame subject to the
Boussinesq approximation for flows with variable density \cite{boussinesq03}.  Our interest is in turbulence that results from
forcing at large horizontal length scales, such as by ocean dynamics much larger than turbulent motions, and so we parametrize the flow in
terms of a velocity scale $\Uhat$ and a length scale $\Lhat$ that are
characteristic of the forcing.  The
remaining quantities used in the parametrization are the acceleration due to
gravity, $\hat{g}$, the ambient density, $\overline{\hat{\rho}}(\hat{z})$, the
molecular viscosity, $\hat{\mu}$, and the thermal diffusivity, $\hat{\alpha}$.
The notation $\hat{\ }$ denotes dimensional quantities which combine to form
the nominal Froude, Prandtl, and Reynolds numbers
\begin{displaymath} 
  \Fl = \frac{2 \pi \Uhat} {\hat{N} \Lhat} , \, \, \, \Pr =
  \frac{\hat{\mu}/\hat{\rho}_0}{\hat{\alpha}} \, \, \, \mbox{and} \, \, \,
  \Rel = \frac{{\Uhat} \Lhat} {\hat{\mu}/\hat{\rho_0}} \, ,
\end{displaymath}
with $\hat{N}^2 = -({\hat{g}}/{\, \, \overline{\hat{\rho}}}) ({d\, \,
  \overline{\hat{\rho}}}/{d\hat{z}})$ the square of the buoyancy frequency and $\hat{\rho}_0$ the reference density.

With the scaling just defined, the dimensionless governing equations are
\begin{subequations}
  \label{eq:goveq}
  \begin{equation}
    \nabla \cdot \vec{u} = 0 \label{eq:cont}
  \end{equation}
  \begin{equation}
    \frac{\partial \vec{u}}{\partial t} + \vec{u} \cdot \nabla \vec{u}
    = - \biggl (\frac{2 \pi}{\Fl} \biggr )^2 \rho \vec{e}_z - \nabla p +
    \frac{1}{\Rel} \nabla^2 \vec{u} + \vec{b} \label{eq:NS}
  \end{equation}
  \begin{equation}
    \frac{\partial \rho}{\partial t} + \vec{u} \cdot \nabla \rho - w
    = \frac{1}{\Rel \Pr} \nabla^2 \rho \label{eq:density} \ .
  \end{equation}
  \label{eq:eqns}
\end{subequations}
Here, $\vec{u} = (u,v,w)$ is the velocity vector, and $\rho$ and $p$ are the
deviations of density and pressure from their ambient values.  The force,
$\vec{b}$, is explained in the next subsection.  Also, $\vec{e}_z$ is the unit
vector in the vertical direction.  The pressure is scaled by the dynamic
pressure, $\hat{\rho}_0 \Uhat^2$, and the density using the ambient density
gradient, i.e., it is scaled by $\Lhat \left| d \overline{\hat{\rho}} / d
\hat{z} \right|$.

We solve the governing equations
numerically using the pseudo-spectral technique described by Riley and de Bruyn Kops \cite{riley03}. Spatial derivatives are computed in Fourier space, the non-linear terms are
computed in real space, and the solution is advanced in time in Fourier space
with the variable-step, third-order, Adams-Bashforth algorithm with pressure
projection.  The non-linear term in the momentum equation is computed in
rotational form, and the advection term in the internal energy equation is
computed in conservation and advective forms on alternate time steps.  These
techniques are standard to eliminate most aliasing errors and ensure conservation of energy, but the simulations
reported in this paper are fully dealiased in accordance with the 2/3 rule via
a spectral cutoff filter.

\subsection{Forcing}
\label{sec:forcing}
The force $\vec{b}$ in \eqref{eq:eqns} is implemented by the deterministic
forcing schema denoted Rf in \cite{rao11}.  The objective is to force all
of the simulations to have the same spectra $E_h\khkz$ with $\kh < \kappa_f$
and $\kz=0$.  The highest wave number forced is $\kappa_f=16\pi / {\cal L}_h$
with ${\cal L}_h$ the horizontal dimension of the numerical domain. 
Deterministic forcing requires choosing a target spectrum
$E_f(\kh<\kappa_f, 0)$.  Unlike for turbulence that is isotropic and
homogeneous in three dimensions, there are no theoretical model spectra for
$E_f$ (c.f.~\cite{overholt97}).  Therefore, run 2 from \cite{lindborg06a}
was rerun using a stochastic forcing schema similar to that used by Lindborg
and denoted schema Qg in \cite{rao11}.  The spectrum for $E_h(\kh<\kappa_f,
0)$ was computed from this simulation and used as the target for the
simulations reported in the current paper. 

In addition to forcing the large horizontal scales, $1\%$ of the forcing
energy is applied stochastically to wave number modes with $\kh=0$ and $\kz =
2\pi j / {\cal L}_v$, $j=2,3,4$.  Here ${\cal L}_v$ is the vertical dimension of
the numerical domain.  This random forcing induces some vertical shear
\cite{lindborg06a}.

\subsection{Dynamically Relevant Variables}
In Portwood et al.\cite{portwood16}, it is assumed that the best variable for identifying the turbulence regimes is $\partial \rho / \partial z$ and that the appropriate metric, shown in Figure~2, is potential enstrophy, which is a particular linear combination of spatial derivatives of $\vec{u}$.  One of the objectives of the research reported here is to verify this assumption by giving the machine learning algorithm access to the basic quantities expected to be relevant based on our understanding flow physics.  By considering \eqref{eq:eqns}, we see that velocities and the spatial derivatives of velocity and density are important. We consider a snapshot of the flow field in time so that temporal derivatives are not relevant for the current analysis but are certainly important dynamically.  The spatial derivatives of pressure respond to acceleration forces to ensure that mass is conserved locally and so, at least for now, we do not consider them to be in the minimum set of quantities to be given to the machine learning algorithm.  Therefore, in the analyses that follow 15 quantities are used:  three velocities components $\vec{u}$, nine velocity derivatives $\nabla \vec{u}$, and three density derivatives $\nabla \rho$.

\section{Data and Pre-processing}

The data consists of data for a snapshot in time of the the flow field.  The simulation domain size has been spectrally filtered to $3842 \times 3842 \times 482$ grid points that are evenly spaced with the smaller dimension in the direction of the gravity vector.  This filtering step applies the 2/3 dealiasing filter in Fourier space that is inherent in the simulation technique as described in the previous section and removes a very small amount of information at the length scales of the smallest turbulent motions in the flow that is important numerically in the simulations but not dynamically for the current purpose.  For this snapshot in time, the 15 quantities defined in the previous subsection (velocity, velocity derivatives, and density derivatives), are stacked to form a 4-dimensional tensor with dimensions
[15, 3842, 3842, 482]. While the simulations use 64-bit arithmetic, for the purposes here the data are reduced to 32-bit floating point with a memory requirement of approximately 426GB. This is
prohibitively large for a proof of concept using a neural network.

To reduce memory requirements and simplify the auto-encoding problem, we
take the first quarter of the extent in both horizontal directions,
giving us a new tensor of shape [15, 961, 961, 482], when 961 = $\lceil
3842 / 4 \rceil$. We then downsample by a factor of 4 in each
spatial direction, giving us a new tensor of shape [15, 241, 241, 121]. This
reduces the total number of elements from $15 \times 3842^2 \times 482 \approx
1.067 \times 10^{11}$ to $15 \times 241^2 \times 121 \approx 1.054 \times
10^8$. Tensors of this size require only $\approx$ 0.42Gb of memory, a much
more manageable size. Thus, it is possible to store approximately 32 examples on a single 16Gb GPU at
a time to use for training. We can repeat the process of downsampling the data to include different
regions of the data set but we do not do that here because our interest is in the reconstruction errors for a single sample.

We normalize the data by subtracting the mean and dividing by the standard deviation on each input channel.  Normalizing by the standard deviation is justified for all input channels because each input variable is approximately Gaussian \cite{debk15}.  The distributions of the longitudinal velocity derivatives are slightly skewed because the flows generate vorticity \cite{taylor38}, but the departures from Gaussian distributions are small for our purposes here.

\section{Results and Discussion}

\subsection{Neural network models}

In a 3D convolutional layer, a number of \textit{filters} are learned that capture features of the input data. That is, $n_f$ weight kernels (three-dimensional tensors) of shape $k_1 \times k_2 \times k_3$ (with channel dimension $d$ = 15, corresponding to the fluid simulation variables) are slid across the input volume with strides ($s_1$, $s_2$, $s_3$). The parameter $k$ is the width, height, and depth of the convolutional filters used to process the input in the convolutional layer, and $s$ is the horizontal, vertical, and depth-wise stride of the filter as it is moved across the input space. For simplicity, we focus on $k = k_1 = k_2 = k_3$ and $s = s_1 = s_2 = s_3$. Convolutional networks exhibit \textit{translation invariance} due to the parameters of convolution kernels being re-used at multiple locations; essentially, the same feature(s) can be recognized in different parts of the input space, and data arranged in a grid (e.g., fluid volumes, images) maintains its spatial structure even as it is processed.

\begin{figure}
    \centering
    \includegraphics[width=0.8\textwidth]{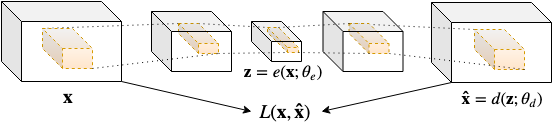}
    \caption{Diagram of convolutional autoencoder architecture with two encoding and two decoding layers. The input data is processed by convolutional kernels that are slid across the input space, reducing its dimensionality. This process is repeated until producing the code tensor ($\mathbf{z}$), which is de-convolved an equal number of times to produce a tensor with the original input size. The input and output tensors are compared via a loss function $L(\mathbf{x}, \mathbf{\hat x})$, and convolution parameters are updated in the negative direction of gradients $\nabla_{\theta_e} L(\mathbf{x}, \mathbf{\hat x})$, $\nabla_{\theta_d} L(\mathbf{x}, \mathbf{\hat x})$.}
    \label{fig:cae}
\end{figure}

We experiment with {convolutional autoencoders} (CAEs) to efficiently process the three spatial dimensions of the data. These are autoencoders which have one or more convolutional layers; in this case, we use three-dimensional convolutions to match the dimensionality of the fluid volume. Since we use only convolutional layers in our method, each element of the code tensor will map to a specific, contiguous region of the fluid volume from which it receives inputs, which can then be analyzed with, e.g., saliency maps to determine how the activation of the code tensor changes as the input data varies \cite{Simonyan2013DeepIC}. A diagram of a convolutional architecture is shown in Figure \ref{fig:cae}.

We experiment with two-, four-, and six-layer CAEs. The two-layer network has a three dimensional convolutional layer with kernel size $k = 3$ and stride $s = 3$, with a variable number of filters $n_f$, and uses the hyperbolic tangent activation function. It maps the input data to a smaller volume, the code tensor. Similarly, for the decoding step, the network has a deconvolutional (transpose convolution) layer with the same kernel size and stride parameters as the convolutional layer, and which maps the code tensor back to a tensor with the same shape as the input data. The four- and six-layer networks have additional convolutional and deconvolutional layers with the same kernel size and stride parameters, which allow us to encode the input data into even smaller code tensors.

The size of the code tensor of the two-layer CAE is $[n_f, 81, 81, 41]$. This has $269,001 \times n_f$ elements; therefore, $n_f < ({1.054 \times 10^8}/{2.69 \times 10^5}) \approx 392$ in order for the code tensor to compress the input volume. We want $n_f$ as small as possible while maintaining a good reconstruction error, in order to maximize compression, making downstream analysis of code tensors more computationally efficient.

\begin{figure}
    \centering \includegraphics[width=\textwidth]{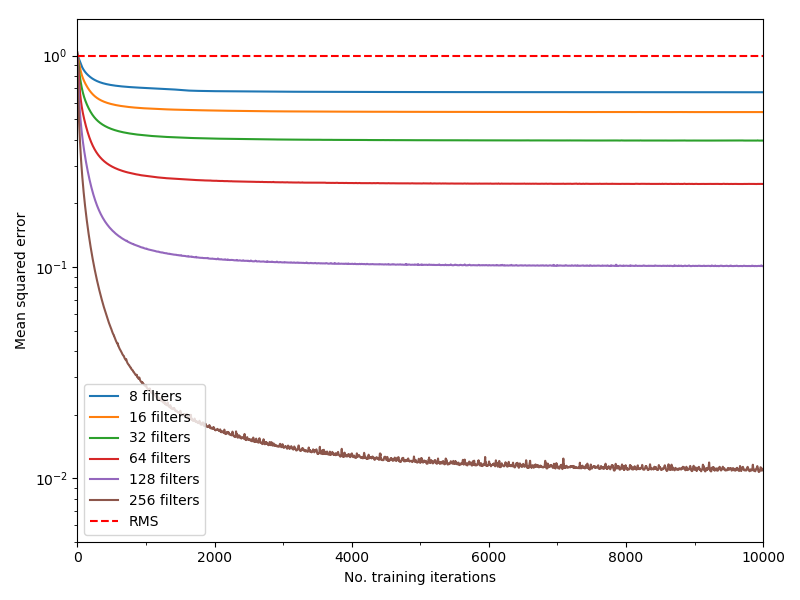}
    \caption{Comparison of loss curves for CAE model \#1 trained for 10K iterations with different settings of $n_{f_1}$. RMS (dotted red line at the top) denotes the root mean square of the input data.  The remaining lines are in the same order, top to bottom, as the legend.}
    \label{fig:loss_curve_comp}
\end{figure}

We vary the number of convolution filters $n_f$ to study the effect on the reconstruction loss from training the two-layer autoencoder on the fluid simulation data. Loss curves for various settings of $n_f$ are compared in Figure \ref{fig:loss_curve_comp}. Using $n_f = 64$, we achieve a compression ratio of approximately $(1.054 \times 10^8) / (64 \times 269,000) \approx 6.5$, with $\approx$17.2M elements in the code tensor and using $\approx$52K learned parameters.

To train the four- and six-layer autoencoders, we adopted a layer-wise training scheme \cite{bengio2006}. For example, to train the four-layer network, we first train a two-layer CAE, then train another, smaller two-layer CAE to reconstruct the former's code tensor. We stack up these two networks and fine-tune the weights of the resulting four-layer network. The size of the four-layer CAE's code tensor is $[n_{f_2}, 27, 27, 15]$, where $n_{f_2}$ is the number of filters in the second convolutional layer. This has $10,935 \times n_{f_2}$ elements. Using $n_{f_2} = 64$, we achieve a compression ratio of $\approx ({1.054 \times 10^8}/{64 \times 10,935}) \approx 151$, with $\approx$700K elements in the code tensor and using $\approx$273K learned parameters.

We trained the six-layer autoencoder using the same stacking procedure as with the four-layer network. The size of the code tensor is $[n_{f_3}, 9, 9, 5]$, where $n_{f_3}$ is the number of filters in the third convolutional layer. This has $405 \times n_{f_3}$ elements. Choosing $n_{f_3} = 64$, we achieve a compression ratio of $\approx 1.054 \times 10^8/(64 \times 405) \approx 4,066$, with $\approx$25.9K elements in the code tensor and using $\approx$495K learned parameters.

\subsection{Reconstruction}

From the data in Table \ref{tab:recon} we can compare the average per-point absolute difference between the input and reconstructed data for a number of architectures and choices of filters. The reconstruction loss generally increases with the depth of the network, and decreases with the size of the code tensor and number of convolution filters.
\begin{table}
    \centering
    \begin{tabular}{||c|c|c|c|c|c||}
        \toprule
        Architecture & $n_{f_1}$ & $n_{f_2}$ & $n_{f_3}$ & Mean abs. diff. & Compression \\
        \midrule
        2-layer CAE & 64 & - & - & 0.3396 & 6.5 \\
        2-layer CAE & 128 & - & - & \textbf{0.2786} & 3.25 \\
        4-layer CAE & 64 & 64 & - & 0.4870 & 151 \\
        4-layer CAE & 64 & 128 & - & 0.4554 & 75.5\\
        4-layer CAE & 128 & 128 & - & 0.4546 & 75.5 \\
        6-layer CAE & 64 & 64 & 64 & 0.5460 & \textbf{4,066} \\
        6-layer CAE & 64 & 64 & 128 & 0.5255 & 2,033 \\
        6-layer CAE & 64 & 128 & 128 & 0.5160 & 2,033 \\
        6-layer CAE & 128 & 128 & 128 & 0.5279 & 2,033 \\
        \bottomrule
    \end{tabular}
    \caption{Comparison of convolutional autoencoder reconstruction loss and compression ratio by architecture and filter size. There is an inherent trade-off between reconstruction error and compression.  The lowest error and the highest compression ratio are in bold.}
    \label{tab:recon}
\end{table}

\begin{figure*}
        \centering
        \begin{subfigure}[t]{0.475\textwidth}
            \centering
            \includegraphics[width=\textwidth]{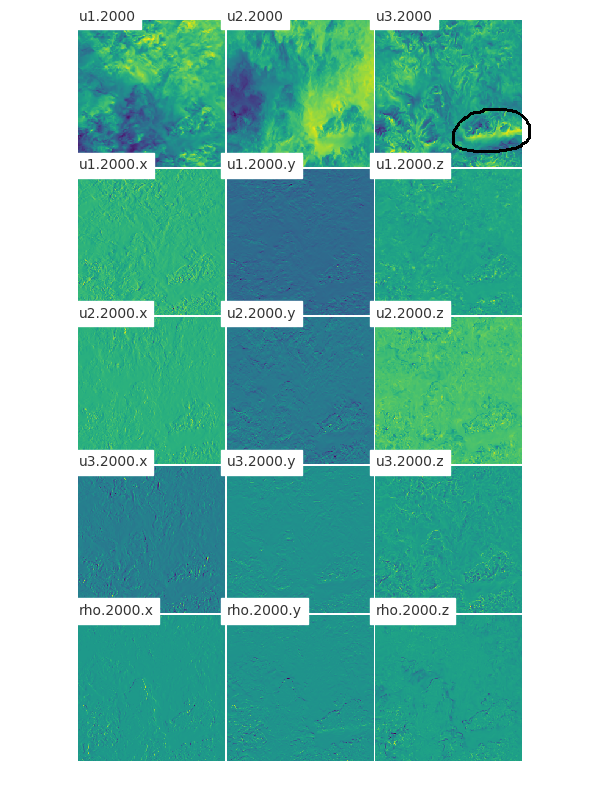}
            \caption{Input data. All 15 fluid variables are shown on a horizontal slice. Circled is an example of a flow feature of interest that appears in the reconstruction, but not the input, of other variables.}
            \label{fig:input}
        \end{subfigure}
        \hfill
        \begin{subfigure}[t]{0.475\textwidth}  
            \centering 
            \includegraphics[width=\textwidth]{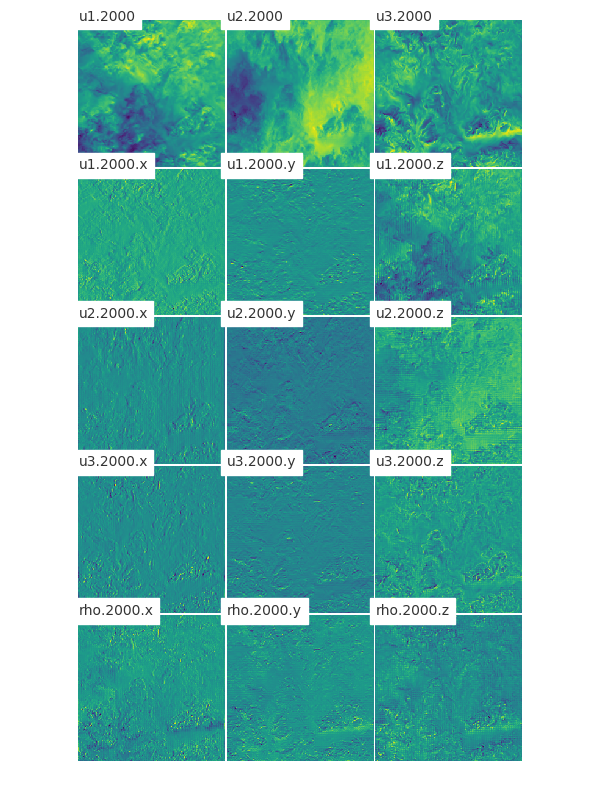}
            \caption{Reconstruction with 2-layer network with $n_{f_1} = 64$.}
            \label{fig:recon1}
        \end{subfigure}
        \vskip\baselineskip
        \begin{subfigure}[t]{0.475\textwidth}   
            \centering 
            \includegraphics[width=\textwidth]{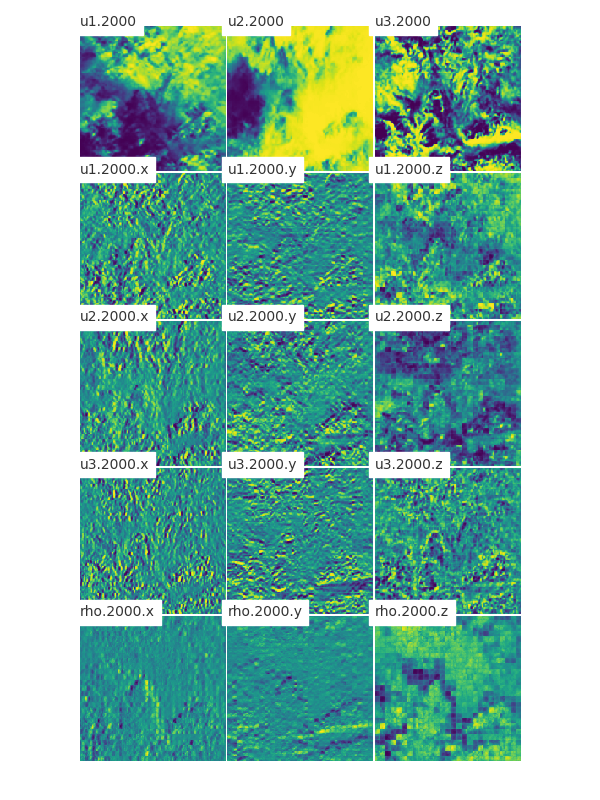}
            \caption{Reconstruction with 4-layer network with $n_{f_1} = n_{f_2} = 64$.}
            \label{fig:recon2}
        \end{subfigure}
        \quad
        \begin{subfigure}[t]{0.475\textwidth}   
            \centering 
            \includegraphics[width=\textwidth]{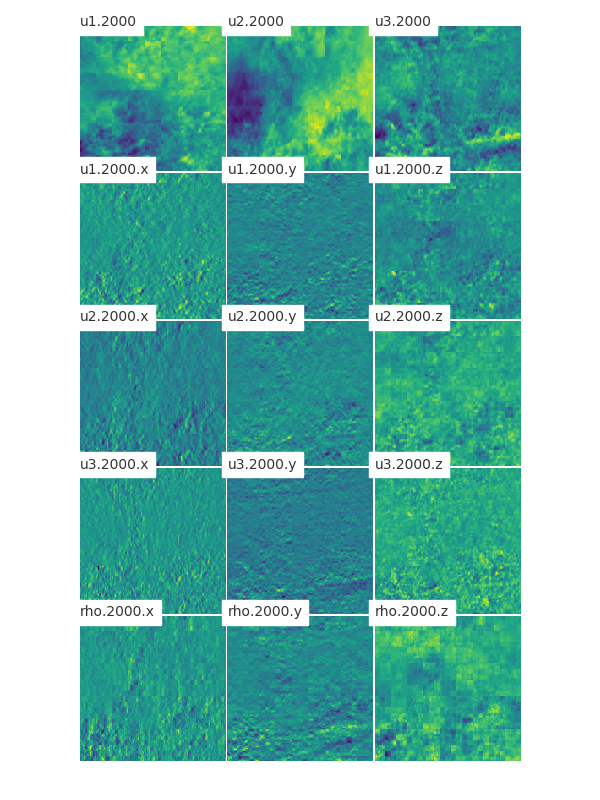}
            \caption{Reconstruction with 6-layer network with $n_{f_1} = n_{f_2} = n_{f_3} = 64$.}
            \label{fig:recon3}
        \end{subfigure}
        \caption{Input data and reconstructions by networks of various depth.}
        \label{fig:recon}
    \end{figure*}

Figure \ref{fig:recon} includes slices viewed from above the input and reconstructed flow fields with the reconstructions done using two-, four-, and six-layer convolutional autoencoders. As the volume of the code tensor decreases, reconstruction loss increases and is visually apparent. We use non-overlapping convolutions, where the kernel size is equal to the stride, so each element in the code tensor is a function of a unique, non-overlapping volume in input space. Similarly, each element in the reconstructed output is a function of a single unique element in the previous convolutional layer. This leads to the grid-like visual artifacts most apparent in Figure \ref{fig:recon3}. These artifacts could be avoided with a different selection of convolutional layer kernel size and stride, wherein neighboring receptive fields overlap, but at the cost of lower compression ratios \cite{odena2016}.

Of interest from the fluid dynamics perspective, some of the input variables appear to `bleed through' in the reconstruction of other input variables. More precisely, since all the quantities in the figure are normalized by the their own standard deviations, the autoencoder causes some information in the flow to be squelched and thereby enhance other features relative to the remainder of the flow field.  In particular, note the circled feature \ref{fig:recon1}, which is not evident in the other inputs but is evident in many of the reconstructions.  Significantly, it is evident in the reconstructions of the 2-, 4-, and 6-layer networks, and is, therefore, robust to the details of the autoencoder. The circled feature is in the input vertical velocity, which has long been recognized as an important identifier of features in SST, c.f.\ figure 9 in \cite{riley03}.  It appears that the autoencoder has identified a key feature of stratified turbulence via a set of mathematical operations applied without information about fluid mechanics.  This result is potentially significant because recall that Portwood et al.\ \cite{portwood16} depended on deep understanding of SST to design their feature recognition algorithm.  A next step in this line of research is to understand why the autoencoder produces results consistent with our understanding of fluid dynamics and then understand what other characteristics of the flow it is identifying of which we were not previously aware.

\subsection{Learned embedding}

Next we consider the embedding of the fluid flow information into a low-dimensional space. In this space, we hope it will be simpler to identify distinct fluid regimes by clustering (or a similar method) which would be otherwise infeasible in the original space. This relies on the assumption that the embedding function preserves useful information about flow regimes contained in the original inputs. To compute a clustering of code tensors, a data set with multiple examples is needed, but here the objective is to understand a snapshot in time of a fluid flow.  

\begin{figure*}
        \centering
        \begin{subfigure}[b]{0.475\textwidth}  
            \centering 
            \includegraphics[width=\textwidth]{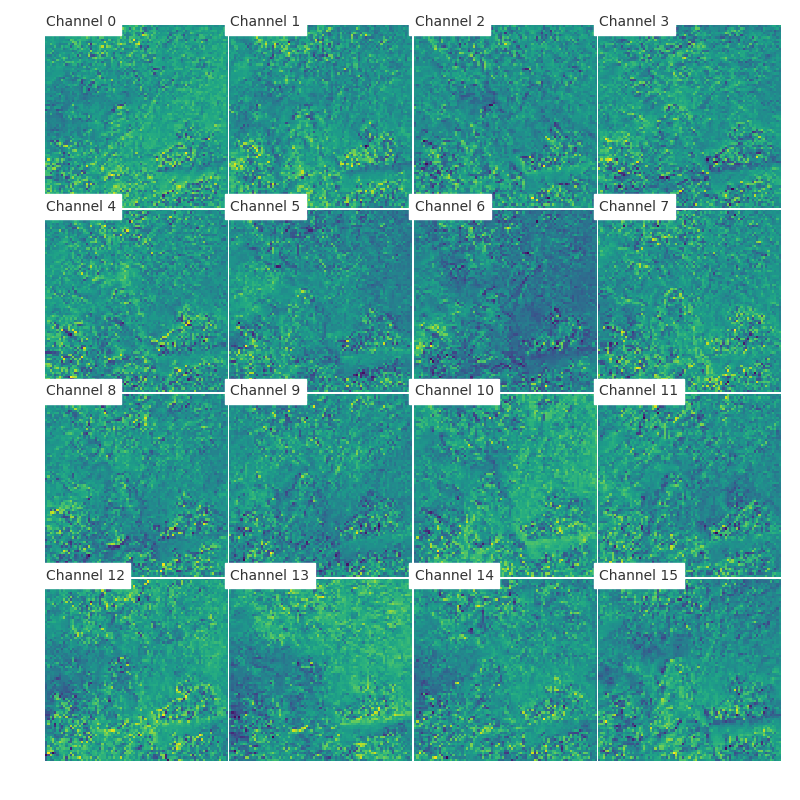}
            \caption{Two-layer CAE code tensor activation with $n_{f_1} = 64$.}
            \label{fig:active1}
        \end{subfigure}
        \vskip\baselineskip
        \begin{subfigure}[b]{0.475\textwidth}   
            \centering 
            \includegraphics[width=\textwidth]{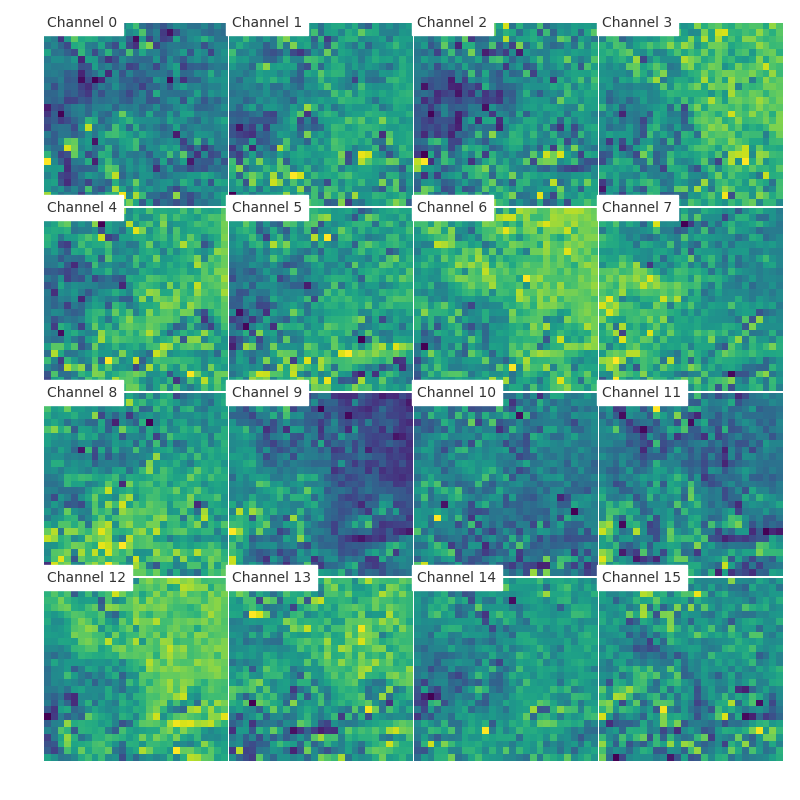}
            \caption{Four-layer CAE code tensor activation with $n_{f_1} = n_{f_2} = 64$.}
            \label{fig:active2}
        \end{subfigure}
        \quad
        \begin{subfigure}[b]{0.475\textwidth}   
            \centering 
            \includegraphics[width=\textwidth]{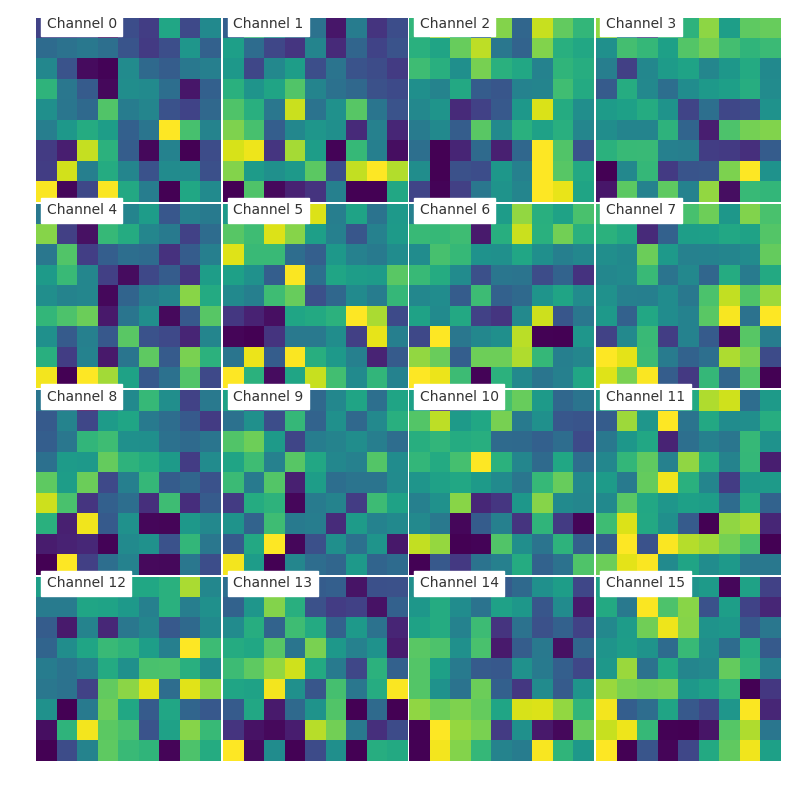}
            \caption{Six-layer CAE code tensor activation with $n_{f_1} = n_{f_2} = n_{f_3} = 64$.}
            \label{fig:active3}
        \end{subfigure}
        \caption{Slices of the code tensor activation by networks of various depth. 64 filters are used by all convolutional layers. 16 randomly sampled channels of the code tensor are shown.}
        \label{fig:activation}
    \end{figure*}

Slices of the code tensor activation for two-, four-, and six-layer networks are shown in Figure \ref{fig:activation}. Code tensor activation is highly non-uniform and appears to capture some of the structure of the input data, especially apparent in the two- and four-layer visualizations; cf. Figure \ref{fig:input}. In particular, the feature circled in Figure \ref{fig:recon} are evident in the two- and four-layer results in Figure \ref{fig:recon1}.  Each filter appears to capture a downsampled version of a combination of one or more of the input variables. This is especially obvious in Figure \ref{fig:recon1}, where, with stride 3, the two-layer CAE appears to have computed a factor of 3 downsampling and learned combination of the input data.

\subsection{Power spectra}
    
\begin{figure}
    \centering
    \includegraphics[width=\textwidth]{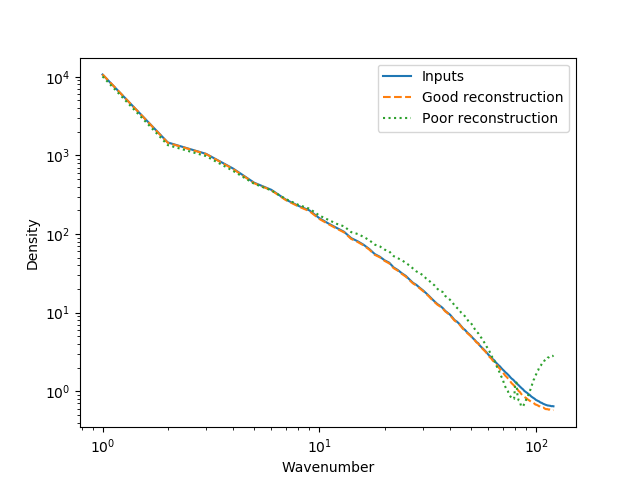}
    \caption{Comparison of the power spectra densities for the input data, a good reconstruction (MSE $\approx$ 0.01), and a poor one (MSE $\approx$ 0.9).}
    \label{fig:power_spectra}
\end{figure}

A standard metric in fluid turbulence is spectra and we compare power spectra densities for the original input data, a ``good'' reconstruction, and a ``poor'' reconstruction, according to the MSE metric, in Figure \ref{fig:power_spectra}. A two-layer CAE with $f_1 = 256$ is used to create the good reconstruction (MSE $\approx$ 0.01), and a two-layer CAE with $f_1 = 8$ is used to create the poor reconstruction (MSE $\approx$ 0.9). The autoencoder with greater capacity and, as a result, lower reconstruction error, is able to match the power spectra density of the inputs quite well, whereas the lower-capacity model struggles to capture higher wavenumbers.  The spectra give us an objective measure of a reconstruction that likely contains too much error for informing us about the fluid flow physics.

Mohan et al.\cite{mohan19} reported that higher wavenumbers are
difficult to reconstruct due to information being lost due to downsampling,
where convolution stride determines the degree of downsampling Increasing stride has the effect of reducing the coverage of a convolution kernel over the input space, with the benefit that the input data is compressed more at lower network depths. Decreasing the stride from 3 to 2 or 1 would likely improve reconstruction at higher wavenumbers at the cost of smaller compression ratios. On the other hand, increasing the stride above 3 would make it longer than the side length of the convolution kernel, thereby causing the kernel to skip over parts of the input data and thus fail to reconstruct even lower wavenumbers. We chose a stride of 3 as a trade-off between compression ratio, network depth, and reconstruction of relatively high wavenumbers given a sufficient number of convolution filters.

\section{Conclusion}
Machine learning is increasingly used to rapidly and objectively identify features in turbulence based on characteristics defined by humans.  In this research, we turn the problem around and ask what machine learning can tell us about a fluid flow that we do not understand.  As a test case, we consider regime identification in stably stratified turbulence (SST).  Using a manual method, Portwood et al.\ \cite{portwood16} identify three turbulence regimes in SST by considering only the vertical derivative of density.  Here we take into account 15 input variables including the three dimensional velocity, its nine spatial derivatives, and the spatial derivatives of density.  The machine learning algorithm is given no cues about fluid mechanics, such as from training samples, beyond that contained in the 15 input variables provided.

The input data considered is a snapshot in time of one of the largest direct numerical simulations of turbulence so that this is a test of machine learning on a realistic data base.  A convolutional autoencoder is used with various numbers of layers to reduce the data and then reconstruct it.  Of interest are the errors in the reconstruction due to information from one input bleeding over to multiple outputs.  It is hypothesized that such bleed over is indicative of an important input variable.  An equivalent interpretation of the results involves considering what characteristics in the inputs are squelched in the outputs as if to indicate that they are not important.  
Indeed, it is observed that features from vertical velocity are prominent in the reconstruction of other output variables and independent of the number of layers in the autoencoder.  The features also appear in the code tensor activation in many of the autoencoder channels.  This is significant because vertical velocity has long be recognized as a marker of key features in SST.  Thus, it appears that the autoencoder, with no information about fluid mechanics other than that contained in the input fields, has reproduced an important piece of human understanding of SST.  The next step in this line of research is to analyze the reconstructions and activation tensors in more detail to teach the fluid dynamicists what other features of the flow the autoencoder is identifying as important or not important.

\section{Acknowledgments}
The simulations used for this research were sponsored by the Office of Naval Research via grant
N00014-15-1-2248.  High performance computing resources were provided through
the U.S.\ Department of Defense High Performance Computing Modernization
Program by the Army Engineer Research and Development Center and the Army
Research Laboratory under Frontier Project FP-CFD-FY14-007.
The research activity of G.D.P.\ is supported by the Department of Energy National Nuclear Security Administration Advanced Simulation and Computing (ASC) program through the Physics and Engineering Models – Mix \& Burn and the Advanced Technology Development and Mitigation projects, and by a Los Alamos National Laboratory Directed Research \& Development project \#20190059DR.

\end{document}